\documentclass{article} 
\usepackage[super,compress]{cite}
\usepackage{amsmath}
\usepackage{amssymb}
\usepackage{graphicx}

\def\calH{{\cal H}}
\def\vev#1{\left\langle #1 \right\rangle}
\def\vevapp#1{\left\langle #1 \right\rangle^{\mathrm{MFPA}}}

\title{Chiral phase transitions in the linear sigma model in the Tsallis nonextensive statistics}
\author{Masamichi Ishihara}

\begin{document}
\maketitle

\begin{abstract}
We studied chiral phase transitions in the Tsallis nonextensive statistics which has two parameters, 
the temperature $T$ and entropic parameter $q$. 
The linear sigma model was used in this study. 
The critical temperature, condensate, masses, and energy density were calculated under the massless free particle approximation. 
The critical temperature decreases as $q$ increases. 
The condensate at $q>1$ is smaller than that at $q=1$.
The sigma mass at $q>1$ is heavier than the mass at $q=1$ at high temperature, 
while the sigma mass at $q>1$ is lighter than the mass at $q=1$ at low temperature. 
The pion mass at $q>1$ is heavier than the mass at $q=1$. 
The energy density increases remarkably as $q$ increases. 
The $q$ dependence in the case of the $q$-expectation value 
is weaker than that in the case of the conventional expectation value with a Tsallis distribution. 
The parameter $q$ should be smaller than $4/3$ from energetic point of view.
The validity of the Tsallis statistics can be determined by the difference in $q$ of the restriction 
for $5/4 < q < 4/3$ when the interaction is weak,
because the parameter $q$ is smaller than $5/4$ in the case of the conventional expectation value with a Tsallis distribution. 
\end{abstract}



\section{Introduction}
Power-like phenomena have been studied in many branches of science, 
and the statistics which describes the phenomena is interested by many researches.
One of them is called Tsallis statistics \cite{TsallisBook} which has been developed past few decades.
The Tsallis statistics is one parameter extension of the Boltzmann-Gibbs (BG) statistics, 
and an introduced parameter $q$ is called entropic parameter. 
A Tsallis distribution approaches a well-known distribution as $q$ goes to $1$.
The Tsallis statistics has been applied to the particle distribution at high energies
\cite{Alberico2000,Biyajima2005,Biyajima2006,Wilk07_Tsallis,Wilk2009,Cleymans2012,Marques2015},  
and it was reported that the distribution is described well by a Tsallis distribution. 
The Tsallis statistics is an possible extension of the BG statistics to describe power-like phenomena.

At high energies, the chiral phase transition is an important topic.
The chiral phase transition is studied by using some models.
The Nambu and Jona-Lasinio (NJL) model and the linear sigma model are often used to investigate the phase transition. 
In such studies, the critical temperature,  the vacuum condensate, and the particle mass are calculated as a function of the temperature.

The phase transition is affected by a power-like distribution. 
The chiral phase transition \cite{Ishihara2015} and associated parametric resonance \cite{Ishihara2016} were studied 
in the linear sigma model in the case of the conventional expectation value with a Tsallis distribution. 
The equation of state in the Tsallis statistics was studied \cite{Drago2004,Pereira2007,Santos2014,Megias2015},
and the phase transition was studied in the NJL model in the Tsallis statistics \cite{Rozynek2009,Rozynek2016}. 
The study of the chiral phase transition in the Tsallis statistics is a significant topic,
because the momentum distribution is described well by a Tsallis distribution. 

The purpose of this paper is to study the effects of the Tsallis statistics on chiral phase transitions by using the linear sigma model.
The $q$ dependence of the critical temperature and that of the energy density are calculated. 
The temperature dependences of the condensate, the sigma mass, and the pion mass, are calculated for various $q$.

The findings are briefly summarized.
The critical temperature decreases as $q$ increases. 
The condensate at $q>1$ is smaller than that at $q=1$.
The sigma mass at $q>1$ is heavier than the mass at $q=1$ at high temperature, 
while the sigma mass at $q>1$ is lighter than the mass at $q=1$ at low temperature. 
The pion mass at $q>1$ is heavier than the mass at $q=1$. 
The energy density increases remarkably as $q$ increases. 
It is found from the calculation of the energy density that the parameter $q$ is smaller than $4/3$.

This paper is organized as follows. 
In Sec.~\ref{sec:eq-of-motion}, 
the expressions of the physical quantities such as critical temperature are given in the Tsallis statistics. 
In Sec.~\ref{sec:numerical-results}, the effects of the Tsallis statistics on physical quantities are shown numerically.
The restriction of $q$ is also shown in the numerical calculations.
Section~\ref{sec:discussion-and-conclusion} is assigned for discussion and conclusion. 


\section{Equations for the condensate, mass, critical temperature, and energy density}
\label{sec:eq-of-motion}
We introduce a function $f_q(\vec{k})$ which is called Tsallis distribution:
\begin{equation}
f_q(\vec{k}) =  \frac{1}{\left[ 1 + (q-1) \beta k \right]_{+}^{1/(q-1)}  + \xi} , 
\qquad (\xi = -1, 0,1) 
, 
\end{equation}
where the quantity $\xi$ is $-1$ for boson, $0$ for classical particle, and $1$ for fermion. 
The function $[x]_{+}$ is $x$ for $x \ge 0$ and $0$ for $x<0$. 
Hereafter, $\xi$ is set to $-1$.

Scalar fields $\phi=(\phi_0, \phi_1, \cdots, \phi_{N-1})$ are used in the linear sigma model,  
and the Hamiltonian density is given by 
\begin{equation}
\calH = \frac{1}{2} \left( \partial^{0} \phi \right)^2 + \frac{1}{2} \left(\nabla \phi \right)^2 + \frac{\lambda}{4} \left( \phi^2 - v^2 \right)^2 - H \phi_0 , 
\label{eqn:Hamiltonian}
\end{equation}
where 
${\displaystyle (\partial^0 \phi)^2 \equiv \sum_{i=0}^{N-1} ( \partial^0 \phi_i) ^2 }$, 
${\displaystyle (\nabla \phi)^2 \equiv \sum_{i=0}^{N-1} ( \nabla \phi_i) ^2 }$, and 
${\displaystyle \phi^2 \equiv \sum_{i=0}^{N-1} \left( \phi_i \right)^2 }$ . 
We divide the field $\phi_i$ as $\phi_i = \phi_{i\rm{c}} + \phi_{i\rm{h}}$, 
where $\phi_{i\rm{c}}$ represents the condensate and $\phi_{i\rm{h}}$ is the remaining part. 
We insert the decomposition into Eq.~\eqref{eqn:Hamiltonian}.

The averaged Hamiltonian density is calculated by taking the average with respect to $\phi_{i \rm{h}}$ under the free particle approximation. 
The expectation value $\vev{\left( \phi_{i\rm{h}} \right) ^{2s+1}} $ is zero, 
where $s$ is a non-negative integer and $\vev{O}$ represents the expectation value of the physical quantity $O$. 
The averaged Hamiltonian density $\vev{\calH(\phi)}$ is given by 
\begin{eqnarray}
\vev{\calH(\phi)} &=& \calH(\phi_{\rm{c}}) 
+ \frac{1}{2} \vev{ \left(\partial^0 \phi_{\rm{h}} \right)^2} + \frac{1}{2} \vev{ \left(\nabla \phi _h\right)^2} 
+  \frac{\lambda}{2} \left(\phi_{\rm{c}}^2 - v^2\right) \vev{\phi_{\rm{h}}^2}  
\nonumber \\  && 
+  \frac{\lambda}{4} \vev{ \left(\phi_{\rm{h}}^2\right)^2}  + \lambda \sum_{j=0}^{N-1} \vev{\phi_{j\rm{h}}^2} \left( \phi_{j\rm{c}} \right)^2 ,
\end{eqnarray}
where ${\displaystyle \phi_{\rm{c}}^2 \equiv \sum_{i=0}^{N-1} \left( \phi_{i\rm{c}} \right) ^2 }$ 
and ${\displaystyle \phi_{\rm{h}}^2 \equiv \sum_{i=0}^{N-1} \left( \phi_{i\rm{h}} \right)^2}$.

The expectation value $\vev{\left( \phi_{j\rm{h}} \right)^2}$ in the massless free particle approximation (MFPA) \cite{Gavin1994,Ishihara1999} 
in the BG statistics is given by 
\begin{equation}
\vevapp{\left( \phi_{j\rm{h}} \right)^2} = \int \ \frac{d\vec{k}}{(2\pi)^3 k} \ f_{q=1}(\vec{k})  ,
\label{expect2:Boltzmann}
\end{equation}
where $f_{q=1}(\vec{k})$ is the distribution function in the BG statistics and the vacuum contribution term is discarded.
The $q$-expectation value in the Tsallis statistics is given by 
replacing $f_{q=1}(\vec{k})$ of Eq.~\eqref{expect2:Boltzmann}  with $\left( f_q(\vec{k}) \right)^q$:
\begin{equation}
\vevapp{\left( \phi_{j\rm{h}} \right)^2} = \int \ \frac{d\vec{k}}{(2\pi)^3 k} \ \left( f_q(\vec{k})  \right)^q . 
\label{expect2:Tsallis}
\end{equation}
For comparison, the conventional expectation value with a Tsallis distribution in MFPA is introduced: 
\begin{equation}
\vevapp{\left( \phi_{j\rm{h}} \right)^2}_{\mathrm{conv}} = \int \ \frac{d\vec{k}}{(2\pi)^3 k} \  f_q(\vec{k}) .
\end{equation}
We define $L_q(T)$ as $L_q(T) := \vevapp{\left( \phi_{j\rm{h}} \right)^2}$ for simplicity, 
where $\vev{\left( \phi_{j\rm{h}} \right)^2}$ is independent of $j$. 
The function $L_q(T)$ is given by 
\begin{equation}
L_q(T) = \frac{1}{2\pi^2} J(\mu=1, \xi=-1; q) , 
\end{equation}
where $J(\mu,\xi; q)$ is defined in \ref{app:sec:integrals}. 
The averaged Hamiltonian density under MFPA, $\vevapp{\calH(\phi)}$, is given by
\begin{eqnarray}
\vevapp{\calH(\phi)} &=& \calH(\phi_{\rm{c}}) 
+  \frac{\lambda}{2} \left[ \left( N+2 \right) \phi_{\rm{c}}^2 - N v^2 \right] L_q(T)
\nonumber \\  && 
+ \frac{1}{2} \vevapp{ \left(\partial^0 \phi_{\rm{h}} \right)^2} + \frac{1}{2} \vevapp{ \left(\nabla \phi _h\right)^2} 
+  \frac{\lambda}{4} \vevapp{ \left(\phi_{\rm{h}}^2\right)^2}  
.
\end{eqnarray}

The potential is tilted to the $\phi_{0\mathrm{c}}$ direction when $H \neq 0$.
Therefore, $\phi_{j\mathrm{c}} $  $(j \neq 0)$ is equal to zero at the minimum of the potential , 
and $\phi_{0\mathrm{c}}$ satisfies the following equation:
\begin{equation}
\left( \phi_{0\mathrm{c}} \right)^3 + \left[ (N+2) L_q(T) - v^2 \right] \phi_{0\mathrm{c}}  - \frac{H}{\lambda} = 0 . 
\end{equation}
The mass squared $\left( m_j \right)^2$ for the field $\phi_j$ is given by 
\begin{equation}
\left( m_j \right)^2 = \lambda \left[ \left( 1+ 2\delta_{j0} \right) \left( \phi_{0\mathrm{c}} \right)^2 + (N+2) L_q(T) - v^2 \right] .
\end{equation}

It is well-known that the critical temperature is not definitely defined when $H \neq 0$. 
In the present paper, the critical temperature $T_{\mathrm{c}}(q)$ is defined as 
the temperature at which a local minimum and a local maximum merge.
The ratio of $T_{\mathrm{c}}(q)$ to $T_{\mathrm{c}}(q=1)$ is given by 
\begin{equation}
\frac{T_{\mathrm{c}}(q) }{T_{\mathrm{c}}(q=1) }
= \sqrt{\Bigg( \frac{\pi^2}{6} \Bigg) \frac{(q-1)}{ \left[\displaystyle \int_0^1 \ ds \ \frac{(s^{1-q} -1)}{(1-s)^q}  \right]  } }
,
\end{equation}
where the critical temperature $T_{\mathrm{c}}(q=1)$ is given by 
\begin{equation}
T_{\mathrm{c}}(q=1) = \sqrt{\frac{12}{(N+2)} \left[ v^2 - \frac{3}{4} \left( \frac{4H}{\lambda} \right)^{\frac{2}{3}} \right] }
.
\end{equation}

The energy density $\varepsilon(q)$ is given by 
\begin{equation}
\varepsilon(q) = \int \ \frac{d\vec{k}}{(2\pi)^3} \ \omega(\vec{k},T)  \left( f_q(\vec{k})  \right)^q ,
\end{equation}
where $\omega(\vec{k},T)$  is the energy of a particle. 
The energy density under MFPA is expressed as 
\begin{equation}
\varepsilon(q) = \frac{1}{2 \pi^2} J(\mu=3,\xi=-1; q) .
\end{equation}
The ratio of $\varepsilon(q)$ to $\varepsilon(q=1)$ is shown in the next section.

\section{Numerical Results}
\label{sec:numerical-results}
In this section, we calculate the physical quantities numerically under MFPA. 
The number of the fields $N$ is set to 4.
The field $\phi_0$ and $\phi_j$ $(j \neq 0)$ are sigma field and pion fields, respectively.
The parameters of the linear sigma model, $\lambda$, $v$, and $H^{1/3}$, are set to 20, 87.4 MeV, and 119 MeV, respectively. 
At zero temperature, these parameters generate $m_{0} = 600$ MeV, $m_{j}=135$ MeV $(j \neq 0)$, and pion decay constant $f_{\pi}=92.5$ MeV. 

\begin{figure}
\begin{center}
\includegraphics[width=0.45\textwidth]{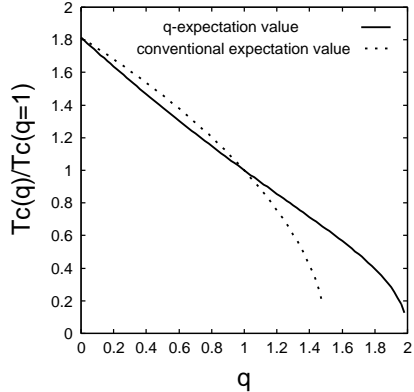}
\end{center}
\caption{The ratio of the critical temperature to the critical temperature at $q=1$.
Solid line is the ratio in the case of the $q$-expectation value, and dashed line is the ratio in the case of the conventional expectation value with a Tsallis distribution. }
\label{Fig:ratio-of-tempertures}
\end{figure}

Figure~\ref{Fig:ratio-of-tempertures} shows 
the ratio of the critical temperature $T_{\mathrm{c}}(q)$ to the critical temperature $T_{\mathrm{c}}(q=1)$ .
The ratio in the case of  the $q$-expectation value is calculated as a function of $q$. 
The ratio in the case of the conventional expectation value with a Tsallis distribution is also calculated, for comparison. 
The ratio decreases as $q$ increases in both the cases. 
The $q$ dependence of the ratio in the case of the $q$-expectation value is weaker 
than that in the case of the conventional expectation value with a Tsallis distribution.
The upper limit of $q$ is restricted in the region of $q<2$ in the case of the $q$-expectation value, as shown in Fig.~\ref{Fig:ratio-of-tempertures}.

\begin{figure}
\begin{center}
\includegraphics[width=0.45\textwidth]{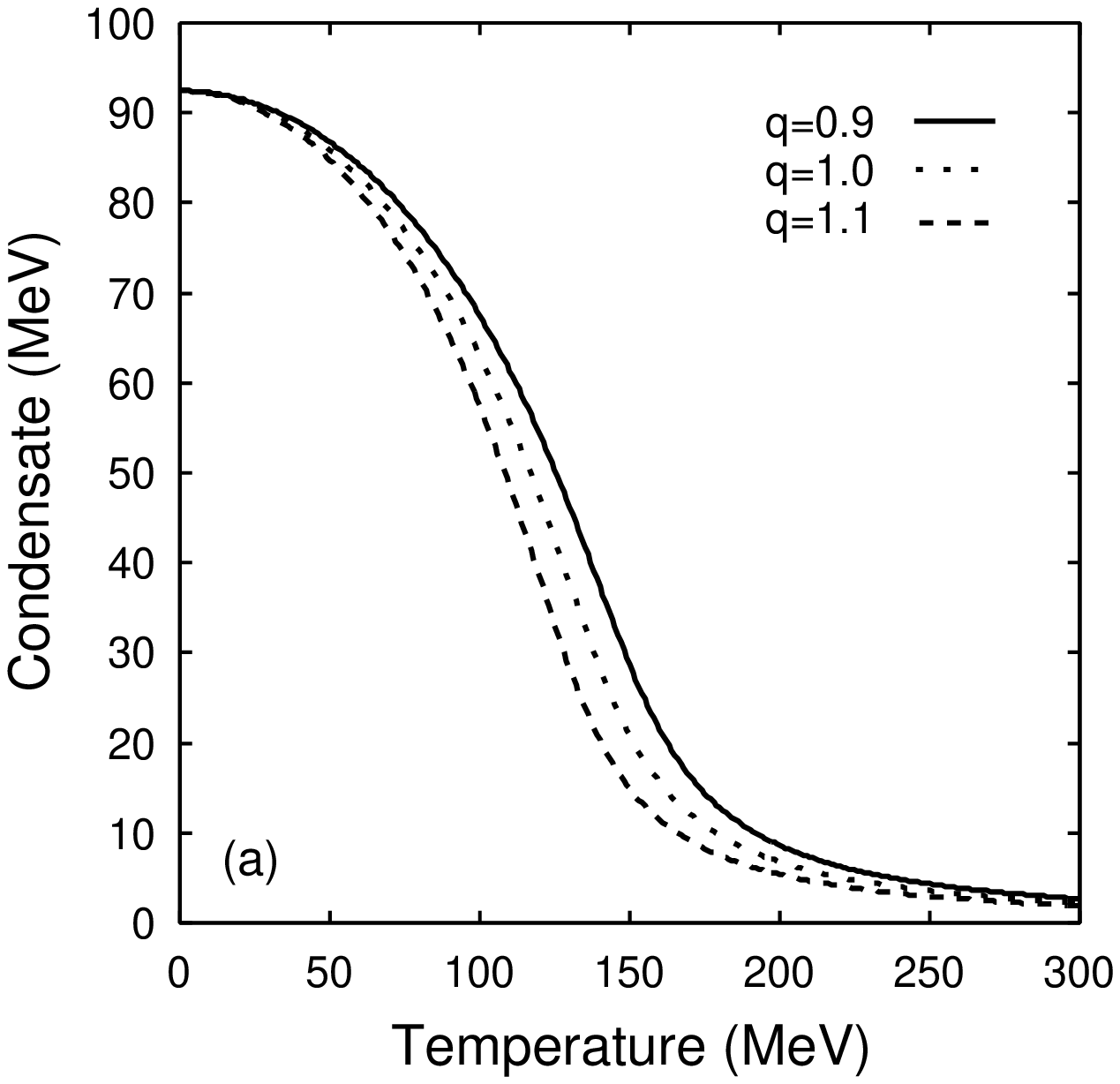}
\includegraphics[width=0.45\textwidth]{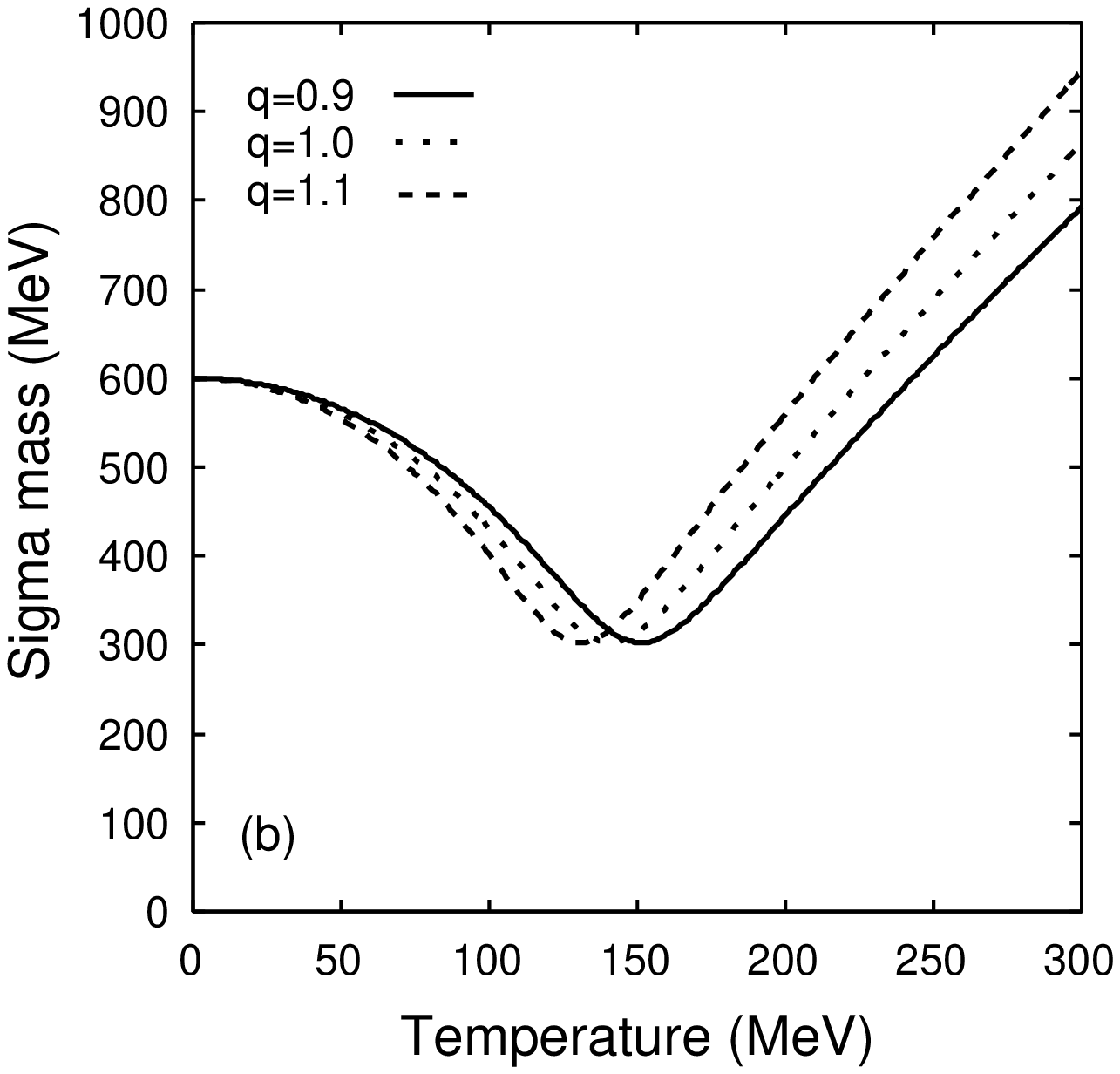}
\includegraphics[width=0.45\textwidth]{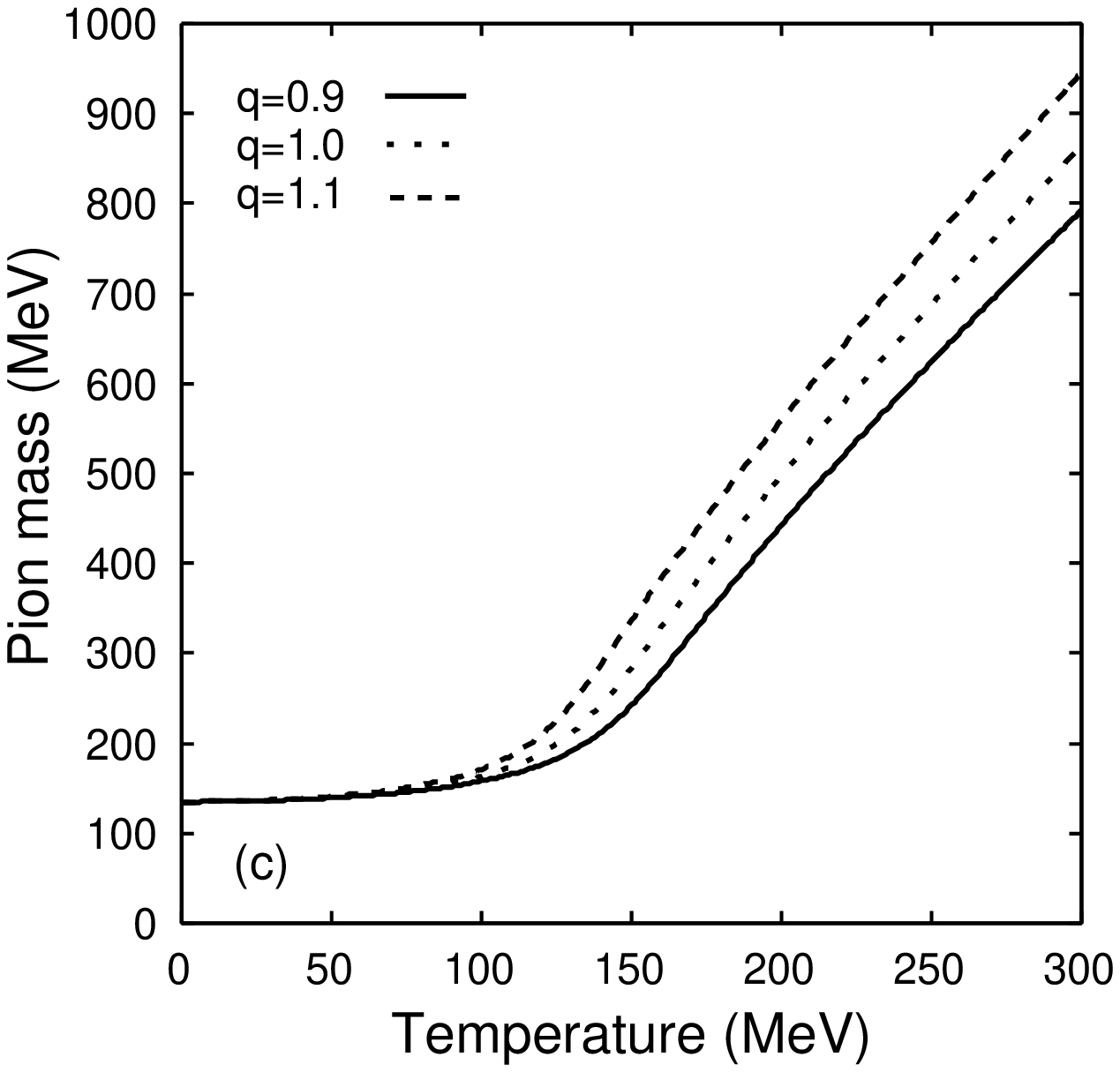}
\end{center}
\caption{Temperature dependence: (a) condensate, (b) sigma mass, and (c) pion mass for $q=0.9$, $1.0$, and $1.1$ in the case of the $q$-expectation value.}
\label{Fig:phi0c-sigmamass-pionmass}
\end{figure}

Figure~\ref{Fig:phi0c-sigmamass-pionmass} shows (a) the temperature dependences of the condensate $\phi_{0\mathrm{c}}$, 
(b) those of the sigma mass $m_0$, and (c) those of the pion mass $m_j$ $(j \neq 0)$ for $q=0.9$, $1.0$, and $1.1$ in the case of the $q$-expectation value.
The condensate at $q>1$ is smaller than that at $q=1$ as shown in Fig.~\ref{Fig:phi0c-sigmamass-pionmass}~(a).
The $q$ dependence of the condensate is reflected in the $q$ dependence of the mass.
As shown in Fig.~\ref{Fig:phi0c-sigmamass-pionmass}~(b), 
the sigma at $q>1$ is  heavier than the mass at $q=1$ at high temperature, 
while the sigma mass at $q>1$ is lighter than the mass at $q=1$ at low temperature. 
The pion mass at $q>1$ is  heavier than the mass at $q=1$.
The $q$ dependences are explained by the $q$ dependence of $L_q(T)$. 
For $q > 1$, the distribution has a long tail,  and the contribution to $L_q(T)$ is large. 
Therefore, the chiral symmetry is restored at low temperature for $q>1$. 

\begin{figure}
\begin{center}
\includegraphics[width=0.45\textwidth]{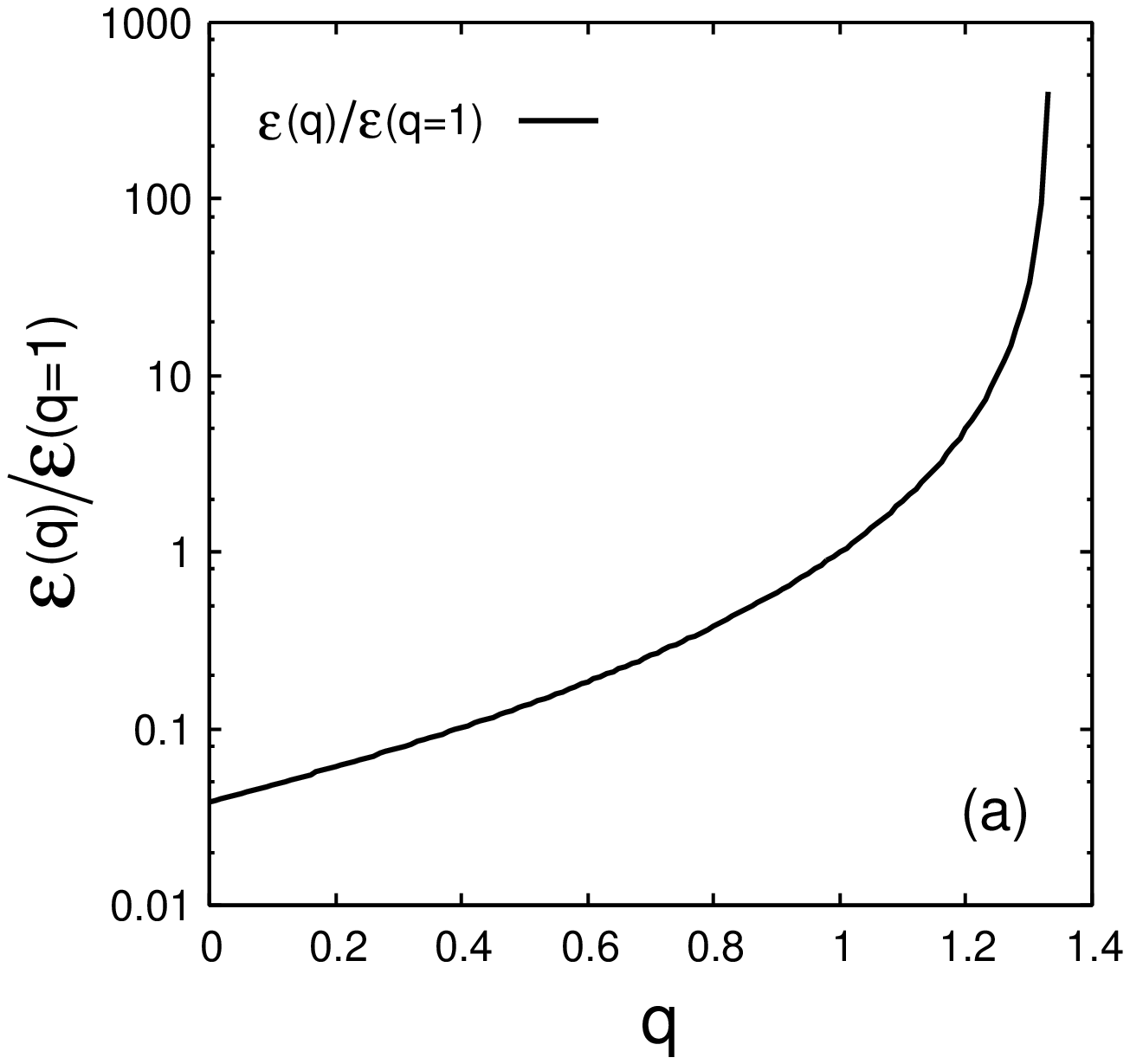}
\includegraphics[width=0.45\textwidth]{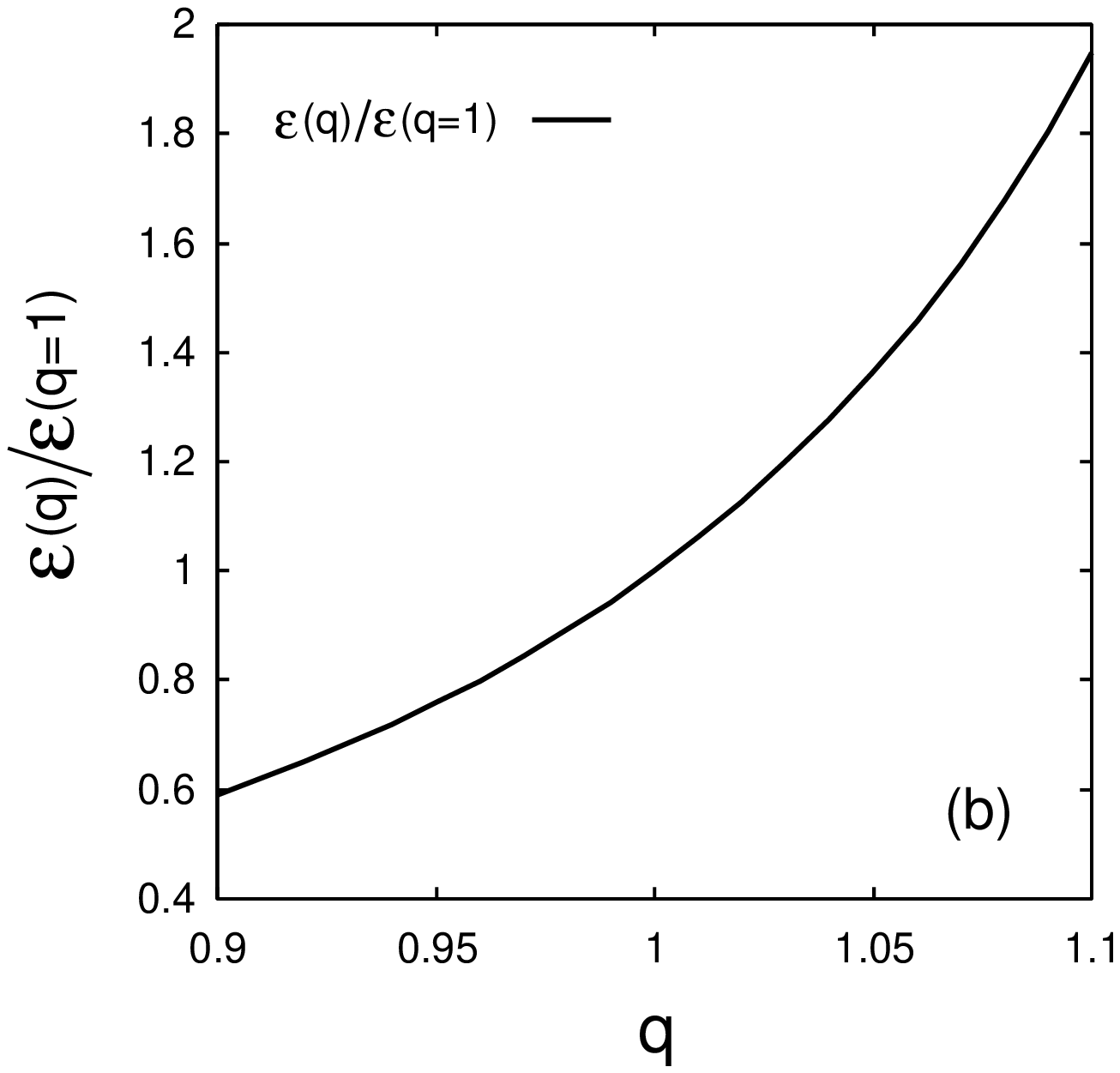}
\end{center}
\caption{The ratio of the energy density, $\varepsilon(q)$, to the energy density at $q=1$, $\varepsilon(q=1)$ 
in the case of the $q$-expectation value:
(a) logarithmic plot in the range of $[0:1.4]$ and  (b) plot in the range of $[0.9:1.1]$.
The value $q$ in Fig.~\ref{Fig:ratio-of-energy-densities}(a) is less than $4/3$. 
}
\label{Fig:ratio-of-energy-densities}
\end{figure}

Figure~\ref{Fig:ratio-of-energy-densities} shows the ratio of the energy density $\varepsilon(q)$ to the energy density 
at $q=1$, $\varepsilon(q=1)$, in the case of the $q$-expectation value.
Figure~\ref{Fig:ratio-of-energy-densities}(a) is the logarithmic plot of $\varepsilon(q)/\varepsilon(q=1)$ in the range of $0 \le q < 4/3$ , 
and Fig.~\ref{Fig:ratio-of-energy-densities}(b) is the plot of $\varepsilon(q)/\varepsilon(q=1)$ in the range of $0.9 \le q \le 1.1$.
The ratio increases remarkably as $q$ increases. 
The restriction of $q<4/3$ comes from the divergence of the integral $J(\mu=3,\xi=-1;q)$ given in \ref{app:sec:integrals}. 
The value of $q$ is smaller than $4/3$ physically, though $q$ is smaller than $2$ in Fig.~\ref{Fig:ratio-of-tempertures}.

\section{Discussion and Conclusion}
\label{sec:discussion-and-conclusion}
We studied the chiral phase transitions in the Tsallis nonextensive statistics.
The linear sigma model was used in the present study. 
There are two parameters, $T$ and $q$,  in the Tsallis statistics, 
where we call the parameter $T$ temperature. 
The critical temperature, condensate, masses, and energy density were calculated in the Tsallis statistics:
the $q$-expectation values of these quantities were calculated.

The critical temperature decreases as $q$ increases. 
The chiral symmetry is restored at low temperature for $q>1$ and at high temperature for $q<1$.
This variation is explained by the fact that a Tsallis distribution for $q>1$ has a long tail. 
This behavior in the case of the $q$-expectation value is the same as that in the case of the conventional expectation value with a Tsallis distribution.
The $q$ dependence of the critical temperature in the case of the $q$-expectation value is weaker than 
that in the case of the conventional expectation value with the Tsallis distribution.

The $q$ dependences of the other quantities are explained in the same manner. 
The condensate at $q>1$ is smaller than that at $q=1$. 
The $q$ dependence comes from the $q$ dependence of $\vevapp{\phi_{j}^2}$, which is affected by the distribution. 
The value $\vevapp{\phi_{j}^2}$ at $q>1$ is larger than that at $q=1$ because of the long tail of the distribution.
Therefore, the condensate decreases as $q$ increases.

The pion mass at $q>1$ is heavier than the mass at $q=1$. 
The $q$ dependence of the pion mass comes from the $q$ dependence of $\vevapp{\phi_{j}^2}$, as discussed above.
The sigma mass at $q>1$ is also heavier than the mass at $q=1$ at high temperature,  
while the sigma mass at $q>1$ is lighter than the mass at $q=1$ at low temperature.
Generally, the sigma mass decreases as $\vevapp{\phi_{j}^2}$ increases, reaches the minimum,  and increases after that. 
The value $\vevapp{\phi_{j}^2}$ at $q>1$ is larger than that at $q=1$ when the temperature $T$ is fixed.
As a result, the sigma mass at $q>1$ is smaller than the mass at $q=1$ at low temperature.
The $q$ dependences of these quantities in the case of the $q$-expectation value are also weaker than 
those in the case of the conventional expectation value with a Tsallis distribution \cite{Ishihara2015}.

The energy density at $q>1$ is extremely larger than that at $q=1$. 
The energy density increases remarkably as $q$ increases. 
This fact implies that the value $q-1$ should be small for $q>1$.
The $q$ dependence of the energy density in the case of the $q$-expectation value is also weaker than 
that in the case of the conventional expectation value with a Tsallis distribution.
The value of $q$ is restricted energetically, and the restriction in the case of the $q$-expectation value is $q<4/3$.
We note that the restriction in the case of the conventional expectation value with a Tsallis distribution is $q<5/4$.

The validity of the Tsallis statistics can be determined by the difference in $q$ of the restriction 
for $5/4 < q < 4/3$ when the interaction is weak. 
The Tsallis statistics will be applied when $q$ is in the range of $5/4<q<4/3$, 
because the value $q$ in the range of $5/4<q<4/3$ is prohibited 
in the case of the conventional expectation value with a Tsallis distribution. 
The restriction of $q$ may give the information of the statistics in various branches related to the nonextensive statistics.

We wish this study is helpful to develop the statistics in order to understand power-like phenomena.


\bibliographystyle{ws-ijmpe}
\bibliography{paper}

\appendix
\section{Integral in the calculations}
\label{app:sec:integrals}
The following integral appears in the calculations:
\begin{align}
J (\mu, \xi; q) &:= \int_0^{k^{*}} \ dk \ k^{\mu} \left( \frac{1}{\left[ 1 + (q-1) \beta k \right]_{+}^{1/(q-1)}  + \xi}\right)^q , 
\\
k^{*} &= 
\left\{
\begin{array}{cc}
\infty&(q > 1)\\
\frac{1}{\beta (1-q)} & (q<1) 
\end{array}
\right.
, 
\nonumber 
\end{align}
where $\mu$ is a positive integer.
The quantity $\xi$ is $-1$ for boson, $0$ for classical particle, and $1$ for fermion. 
The notation $[x]_{+}$ is defined by 
\begin{equation}
[x]_{+} = \left\{ \begin{array}{cl} x &\qquad  (x \ge 0)\\ 0 &\qquad  (x < 0) \end{array} \right.
.
\end{equation}
This integral is rewritten by changing of variables:
\begin{equation}
J (\mu, \xi; q) = \frac{1}{\beta^{\mu+1} (q-1)^{\mu}} \int_{0}^{1} \ ds \ \frac{(s^{1-q} -1)^{\mu}}{(1+\xi s)^q} 
.
\end{equation}

The integral $J(\mu,\xi=-1;q)$ appears in the present calculations. 
This integral at $q>1$ converges when ${\displaystyle q  < 1 + 1/\mu}$. 
The integral can be numerically evaluated for ${\displaystyle q<1+1/\mu}$.
The integral  for $q < 1$ is represented with Beta function $B(x,y)$:
\begin{align}
&
J(\mu, \xi=-1; {q<1})  
\nonumber \\  &\qquad 
= \frac{1}{\beta^{\mu+1} (q-1)^{\mu}}  
\sum_{n=0}^{\mu} \left( \begin{array}{c} \mu \\ n \end{array} \right) (-1)^{\mu - n} B\left( n(1-q)+1,1-q \right) 
. 
\end{align}
The integral  $J(\mu,\xi=-1;q=1)$ is simply represented as
\begin{equation}
J(\mu, \xi=-1; {q=1})  = \frac{\mu !}{\beta^{\mu+1}} \sum_{r=1}^{\infty} \frac{1}{r^{\mu+1}} 
.
\end{equation}

In the last of this appendix, we give the expression of the integral for classical particle ($\xi=0$).
For $q<1$ and $1<q<1+1/\mu$, the integral is represented as
\begin{subequations}
\begin{align}
J (\mu, \xi=0; q) 
=\frac{1}{\beta^{\mu+1} (1-q)^{\mu}} \sum_{n=0}^{\mu} \left( \begin{array}{c} \mu \\ n \end{array} \right) (-1)^{n} \left( \frac{1}{n(1-q) + 1} \right) . 
\end{align}
The integral $J (\mu, \xi=0; q=1)$ is given explicitly by 
\begin{align}
&  J (\mu, \xi=0; q=1) =  \frac{\mu !}{\beta^{\mu+1}} . 
\end{align}
\end{subequations}


\end{document}